\documentclass[aps,twocolumn,superscriptaddress,showpacs]{revtex4}
\usepackage{feynmp}
\usepackage{amssymb}
\usepackage{graphicx}
\usepackage{subfigure}
\usepackage{color}

\usepackage{hyperref}

\begin{document}

\title{Probing the Topological Phase Transition via Density Oscillations
in Silicene and Germanene}

\author{Hao-Ran Chang}
\affiliation{Department of Physics and Institute of Solid State Physics, Sichuan Normal University, Chengdu, Sichuan 610066, China}

\author{Jianhui Zhou}
\email{jhzhou@andrew.cmu.edu}
\affiliation{Department of Physics, Carnegie Mellon University, Pittsburgh, Pennsylvania 15213, USA}

\author{Hui Zhang}
\affiliation{Department of Physics and Astronomy, Ohio University, Athens, Ohio 45701, USA}

\author{Yugui Yao}
\affiliation{School of Physics, Beijing Institute of Technology, Beijing 100081, China}


\begin{abstract}
We theoretically investigated two kinds of density oscillations, the
Friedel oscillation and collective excitation in the silicene and
germanene within the random phase approximation, and found that the tunable
spin-valley coupled band structure could lead to some exotic properties
in these two phenomena. Based on an exact analytical and numerical analysis,
we demonstrated that the beating of the screened potential as well as the
undamped plasmon mode can be taken as fingerprints of a topological
phase transition in doped silicene and doped germanene. Thus our proposal
here establishes the connection between the topological phase transition
and the density oscillations that can be accessed by a variety of
experimental techniques.
\end{abstract}


\maketitle

The quantum spin Hall effect (QSHE) \cite{KaneMele,BHZ} has been
studied extensively in both theoretical and experimental aspects.
It is well known that topologically protected helical edge states
are a distinct feature which characterizes the QSHE \cite{Molenkamp}.
The transport measurement of edge states requires that the bulk state
must be insulating. In practice, the system is usually metallic resulting
from defects, self-doping and charge transfer from metallic substrates, and
thus the detection of the topological phase transition (TPT) when the bulk
is metallic becomes urgent and important in two-dimensional systems. In
this Rapid Communication, we connect the TPT with two kinds of density oscillations: Friedel oscillation \cite{Friedel} and collective excitation in silicene and germanene.

Silicene \cite{Voon07,Cahangirov}, a single layer of silicon atoms forming
a two-dimensional (2D) buckled honeycomb lattice, can be regarded as the
silicon-based counterpart of graphene \cite{Si2DGe}. The buckled honeycomb
structure gives rise to a tunable spin-valley coupled band structure, which
accounts for many exotic transport and superconducting phenomena
\cite{Ezawa,WFTsai,HaizhouLu,PanPRL,FLiu,WHWan,LDZhang} and makes
silicene a promising candidate for the QSHE \cite{PRLYao}. So far
silicene or its superstructure has only been synthesized on
metallic surfaces \cite{Vogt,LChenPRL109,LeiMeng,Fleurence}, hence
the transport measurement of the helical edge states is prevented
due to the metallic bulk state. On the other hand, it has been claimed
that a Dirac-like spectrum does exist in silicene from experimental
observations \cite{LChenPRL109,Vogt}. Although there is some debate
about the origin of the linear dispersion \cite{HaiPPeng}, it is still
highly worthwhile to examine whether or not silicene hosts the QSH state.

\begin{figure}[htbp]
\centering
\subfigure{\includegraphics[width=4cm]{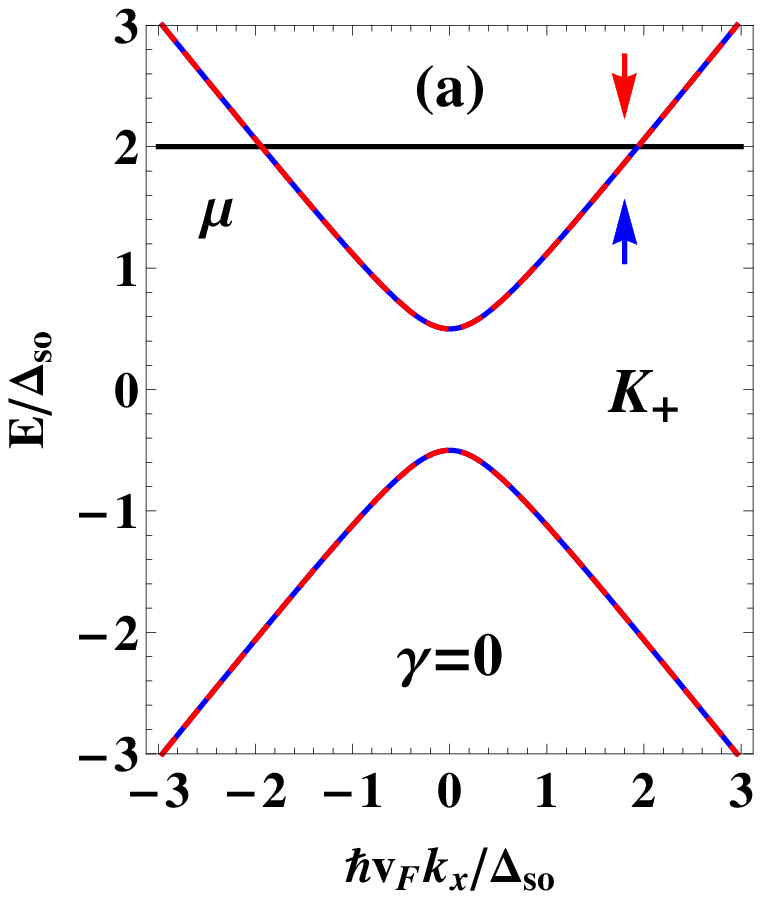}}
\subfigure{\includegraphics[width=4cm]{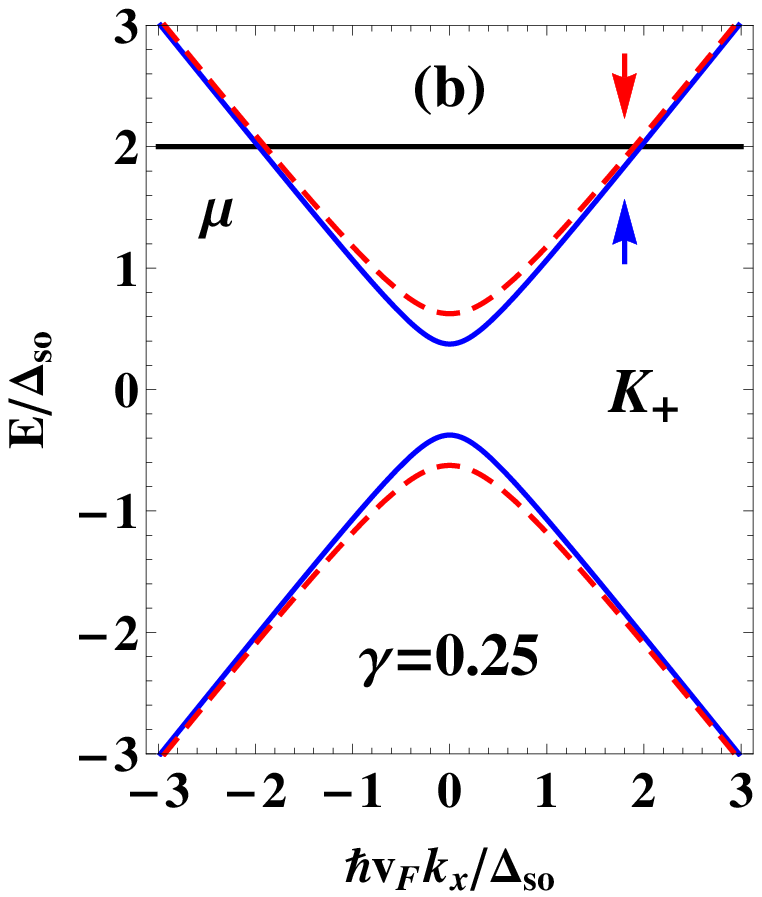}}\\
\subfigure{\includegraphics[width=4cm]{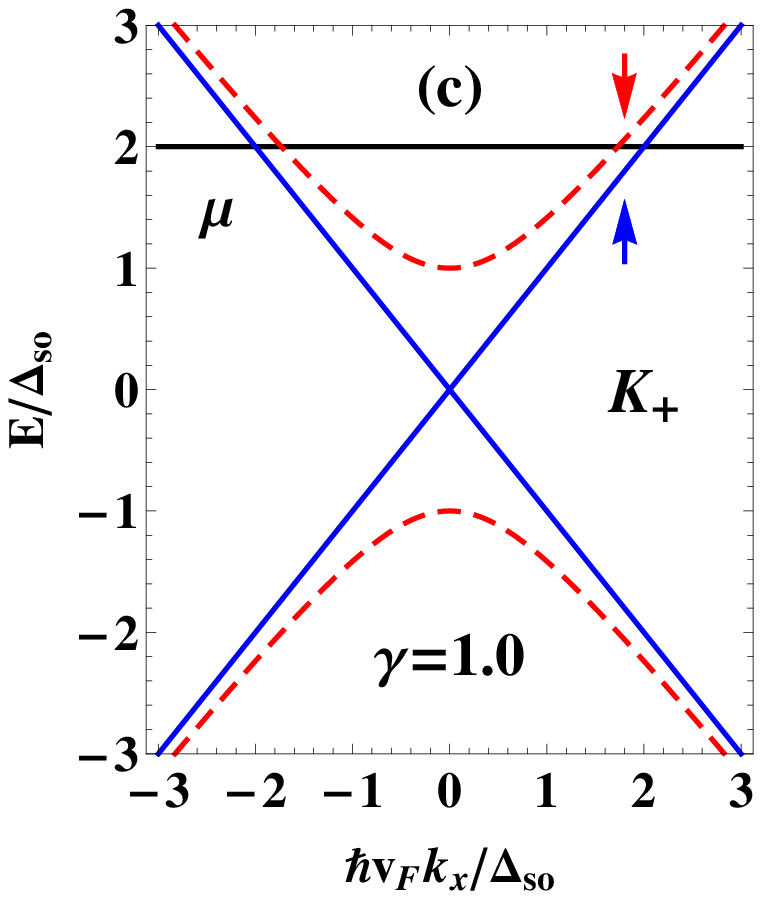}}
\subfigure{\includegraphics[width=4cm]{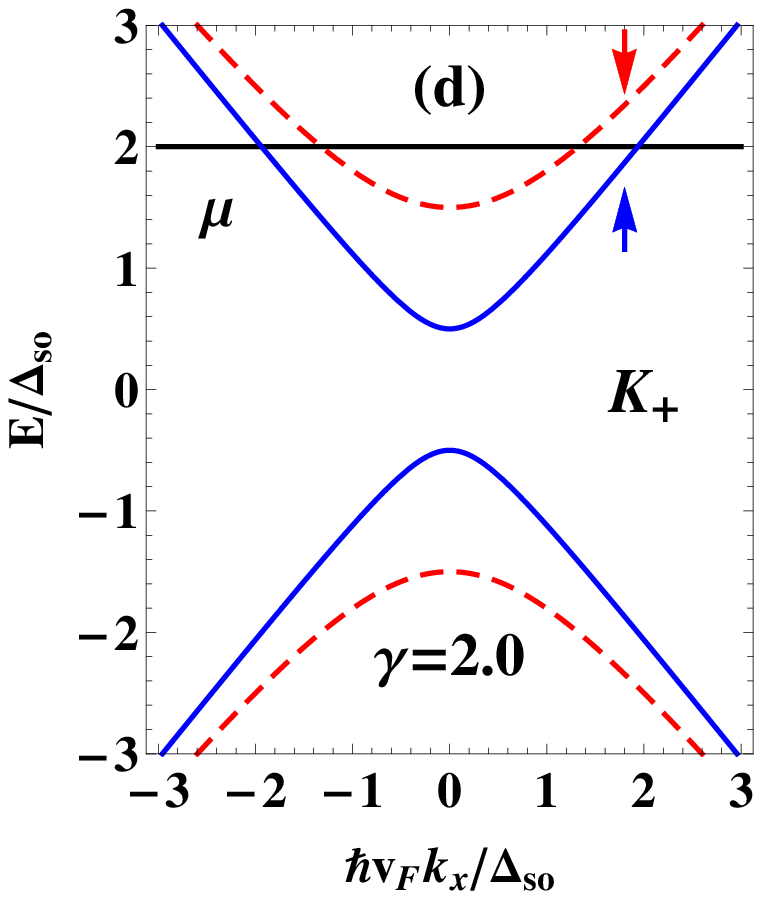}}\\
\caption{(Color online) The band structures at the $K_{+}$
valley for different $\gamma$'s. The solid blue line
represents the spin-up band while the dashed red line
corresponds to the spin-down band.}
\label{Fig1}
\end{figure}

In this Rapid Communication, we propose a detection method of TPT by employing
both the Friedel oscillation and collective excitation in silicene,
and show how to extract the information about the TPT from these two effects.
First, the screened potential of charged impurities has a beating
structure of Friedel oscillations (the interference pattern of two
branches of density waves of electrons). As one band gap decreases
(for example, the spin-up gap in Fig. \ref{Fig1}), the beating of
the screened potential gradually becomes faint and eventually
vanishes at the TPT point. Second, the undamped plasmon mode (UPM)
that emerges in the single-particle excitation (SPE) gap disappears
when approaching the TPT point and reappears after it. Therefore,
these two kinds of density oscillations can be used to detect the
TPT in metallic silicene and germanene.

We start from a simple low-energy effective Hamiltonian in silicene
under an external electric field\cite{PRLYao,YaoPRB},
\begin{equation}
\hat{H}_{\xi}=\hbar v_{F}\left(\xi k_{x}\tau_{x}+k_{y}\tau_{y}\right)
-\xi\Delta_{\mathrm{so}}\sigma_{z}\tau_{z}/2+\Delta_{z}\tau_{z}/2,
\label{hamiltonian}
\end{equation}
where the subscripts $\xi=\pm$ denote the two inequivalent valleys or
Dirac points $K_{+}$ and $K_{-}$, respectively. The first term describes
the behavior of 2D Dirac electrons with the Fermi velocity $v_{F}$.
The second term is the intrinsic spin orbit coupling (SOC) term with
a magnitude of $\Delta_{\mathrm{so}}$. The last term is due to the $A-B$
sublattice symmetry breaking and is defined as $\Delta_{z}=E_{z}\cdot d$
, where $E_{z}$ is the effective external electric field perpendicular
to the sample including all of the screening effect, and $d$ is the perpendicular
distance between the two sublattice planes. Pauli matrices $\tau_{i}$
and $\sigma_{i}$ act on the pseudospin space related to the $A$ and
$B$ sublattices and the real spin degree of freedom, respectively. Since the strength of the intrinsic Rashba SOC is much smaller than that of the intrinsic SOC \cite{YaoPRB}, we can neglect it here. Thus, the effective Hamiltonian can be
classified by the eigenvalues of $\sigma_{z}$ and the corresponding
Hamiltonian of the $K_{+}$ valley can be expressed as
\begin{equation}
\hat{h}_{\sigma}=\left(\begin{array}{cc}
\left(\Delta_{z}-\sigma\Delta_{\mathrm{so}}\right)/2&
\hbar v_{F}\left(k_{x}-ik_{y}\right)\\
\hbar v_{F}\left(k_{x}+ik_{y}\right)& -\left(\Delta_{z}-\sigma\Delta_{\mathrm{so}}\right)/2
\end{array}\right),\label{HamSpin}
\end{equation}
where $\sigma=\pm$ refer to the spin-up and spin-down bands, respectively.
Two remarks are in order. First, the corresponding Hamiltonian of
the $K_{-}$ valley can be obtained from Eq.$\left(\ref{HamSpin}\right)$
via the time reversal operation. Second, we assume that there are
no short-range impurities and defects that cause intervalley scattering.
Accordingly, we are able to restrict our discussion to the case of
a single valley, and then multiply the valley-degeneracy factor $g_{v}=2$
to the final results. The eigenvalues of the effective Hamiltonian in Eq.$\left(\ref{HamSpin}\right)$ can be evaluated straightforwardly as $E_{\sigma\lambda}=\lambda\sqrt{\hbar^{2}v_{F}^{2}k^{2}+\Delta_{\sigma}^{2}}$,
where $\lambda=\pm$ are for the conduction and valence bands, and
$2\Delta_{\pm}=\Delta_{\mathrm{so}}\left|\gamma\mp1\right|$ are the
spin dependent energy gaps with $\gamma=\Delta_{z}/\Delta_{\mathrm{so}}$.

Before presenting the detailed calculations, we first discuss the topological
phases of our model Hamiltonian. The model would describe the gapless graphene
for $\Delta_{\mathrm{so}}=0$ and gapped graphene for $\Delta_{\mathrm{so}}\neq0$
and $\gamma=0$, [see Fig. \ref{Fig1}$(\mathrm{a})$]. When $\Delta_{\mathrm{so}}\neq0$
and $\gamma\neq0$, the model will account for silicene. The region for $|\gamma|<1$
pertains to the QSH state with a pair of helical edge states, as shown in Fig. \ref{Fig1}$(\mathrm{b})$, and for $|\gamma|>1$ the system is the usual band insulator without any topological edge states [see Fig. \ref{Fig1}$(\mathrm{d})$]. Interestingly, when $\gamma=1$, i.e., $\Delta_{z}=\Delta_{\mathrm{so}}$, at the $K_{+}$ valley the spin-up band becomes gapless and the gap of the spin-down band is $\Delta_{\mathrm{so}}$, as shown in Fig. \ref{Fig1}$(\mathrm{c})$, while at the $K_{-}$
valley the spin reverses, which has been termed a valley-spin-polarized semimetal \cite{Ezawa} and is a unique characteristic of silicene.

In the following we want to investigate the Friedel oscillation and
the plasmon spectrum. The analytic dielectric function within the
random phase approximation (RPA) is given by
\begin{equation}
\varepsilon(q,\omega)=1-V(q)\chi_{0}(q,\omega),\label{dielFunction}
\end{equation}
where $V\left(q\right)=2\pi e^{2}/\kappa q$ is the Fourier
transform of the 2D Coulomb interaction, $V\left(r\right)=e^{2}/\kappa r$,
and $\kappa$ is the effective background dielectric constant \cite{RelDieConst}.
The full formal expression of the polarization function in silicene is given
by \cite{KWKShung}
\begin{eqnarray}
&&\chi_{0}\left(q,\omega\right)
=g_{v}\sum_{\lambda,\lambda^{\prime},\sigma=\pm}
\int\frac{d^{2}k}{\left(2\pi\right)^{2}}\left|\left\langle \psi_{\sigma\lambda}\left({\bf k}\right)|\psi_{\sigma\lambda^{\prime}}
\left({\bf {\bf k}^{\prime}}\right)\right\rangle \right|^{2}\nonumber \\
&&\hspace{2.5cm}\times\frac{n_{F}\left[E_{\sigma\lambda}\left({\bf k}\right)\right]-n_{F}\left[E_{\sigma\lambda^{\prime}}\left({\bf {\bf k}^{\prime}}\right)\right]}{\hbar\omega+E_{\sigma\lambda}\left({\bf k}\right)-E_{\sigma\lambda^{\prime}}\left({\bf {\bf k}^{\prime}}\right)+i\eta},\label{chi0QOmega}
\end{eqnarray}
where $\left|\psi_{\sigma\lambda}\left({\bf k}\right)\right\rangle$
is the eigenstate corresponding to $E_{\sigma\lambda}$, $\eta$ is
a positive infinitesimal quantity, ${\bf k}^{\prime}={\bf k+q}$, and $n_{F}(x)=\left[\exp\left\{\beta\left(x-\mu\right)\right\} +1\right]^{-1}$,
with $\beta=1/k_{B}T$. Noted that since the spin-up component and
spin-down component decouple with each other, the total polarization
function turns out to be the summation of those for both spin
species. The polarization function for a certain spin species in silicene
is mathematically identical to the case of a gapped graphene.

To proceed with the theoretical details, we assume zero temperature
$T=0\:\mathrm{K}$, and then the noninteracting Fermi function $n_{F}\left(x\right)$
turns into a simple step function $\theta(\mu-x)$. We restrict our
discussion to positive frequencies $\omega>0$ because of the general
relation $\chi_{0}(q,-\omega)=[\chi_{0}(q,\omega)]^{\ast}$. Due to
the electron-hole symmetry, the plasmon in both $n$- and $p$-doped
samples would show the same dynamical behaviors. Therefore, we can
concentrate on the $n$- doped case, namely, the finite chemical potential
$\mu$ lies in the conduction band. Since when $\Delta_{+}\le\mu\le\Delta_{-}$
there is no beating in the screened potential, in our discussion
of Friedel oscillation, we require that the Fermi contour consists
of two Fermi circles with different Fermi wave vectors, $k_{F}^{\sigma}=\sqrt{\mu^{2}-\Delta_{\sigma}^{2}}/\hbar v_{F}$
as long as the external electric field changes.

The static screened potential of a charged impurity is given by the
integration of the dielectric function,
\begin{equation}
\phi(r)=\frac{Ze}{\kappa}
\int_{0}^{\infty}dq\frac{J_{0}(qr)}{\epsilon(q,0)},\label{phi}
\end{equation}
where $\varepsilon(q,0)=1-V(q)\chi_{0}(q,0)$ is the static dielectric
function, $Ze$ the charge of the impurity and $J_{0}(x)$ the zeroth
order Bessel function of the first kind, which comes from the integration
over the angular variable. After some cumbersome manipulations, one can
obtain the static wave vector dependent polarization function from
Eq.(\ref{chi0QOmega}),
\begin{eqnarray}
&&\chi_{0}(q,0) =-\frac{2\mu}{\pi\hbar^2v_{F}^2}\Big[1-\sum_{\sigma=\pm}\theta(q-2k_{F}^{\sigma})
\Big(\frac{\sqrt{q^{2}-4k_{F}^{\sigma2}}}{4q}\nonumber\\
&&\hspace{1.0cm}-\frac{\hbar^2v_{F}^2q^{2}-4\Delta_{\sigma}^{2}}{8\hbar v_{F}q\mu}
\arctan\frac{\hbar v_{F}\sqrt{q^{2}-4k_{F}^{\sigma2}}}{2\mu}\Big)\Big].\label{staticchi}
\end{eqnarray}

\begin{figure}[htbp]
\centering
\subfigure{\includegraphics[width=4cm]{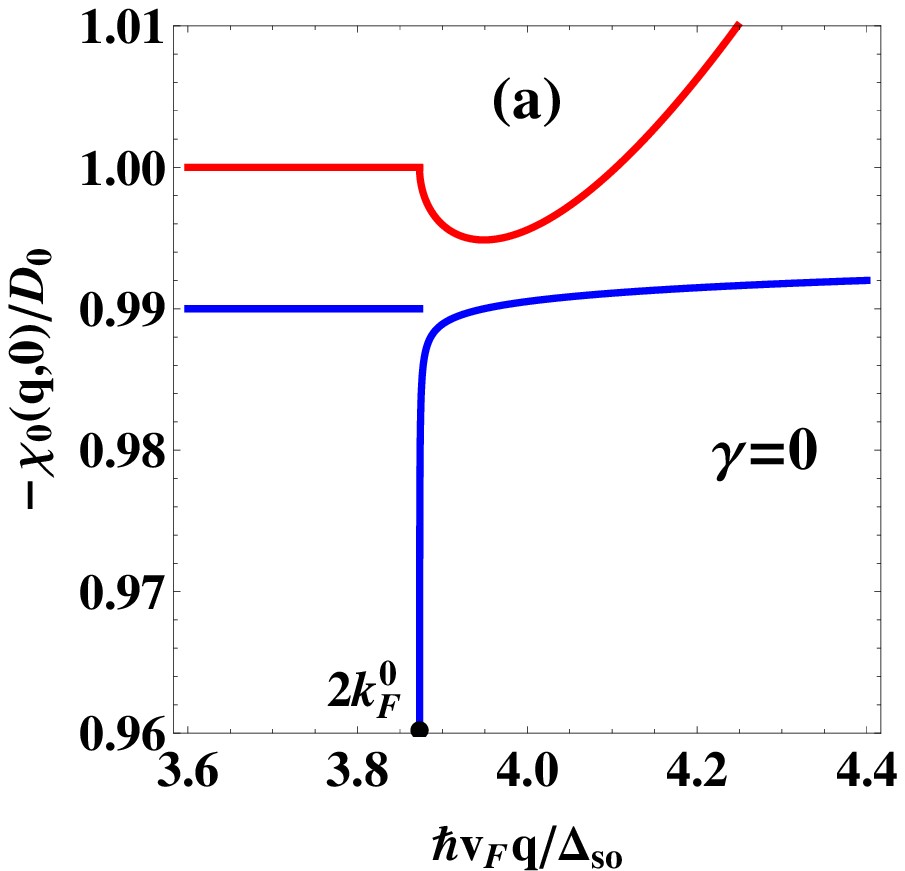}}
\subfigure{\includegraphics[width=4cm]{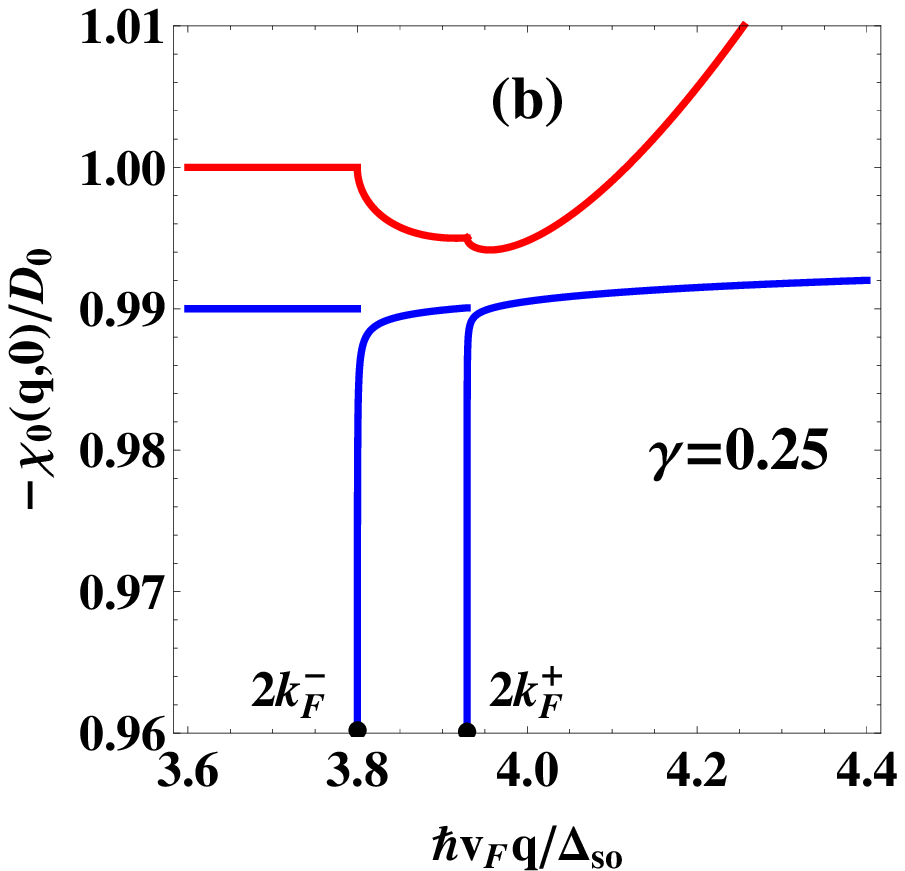}}\\
\subfigure{\includegraphics[width=4cm]{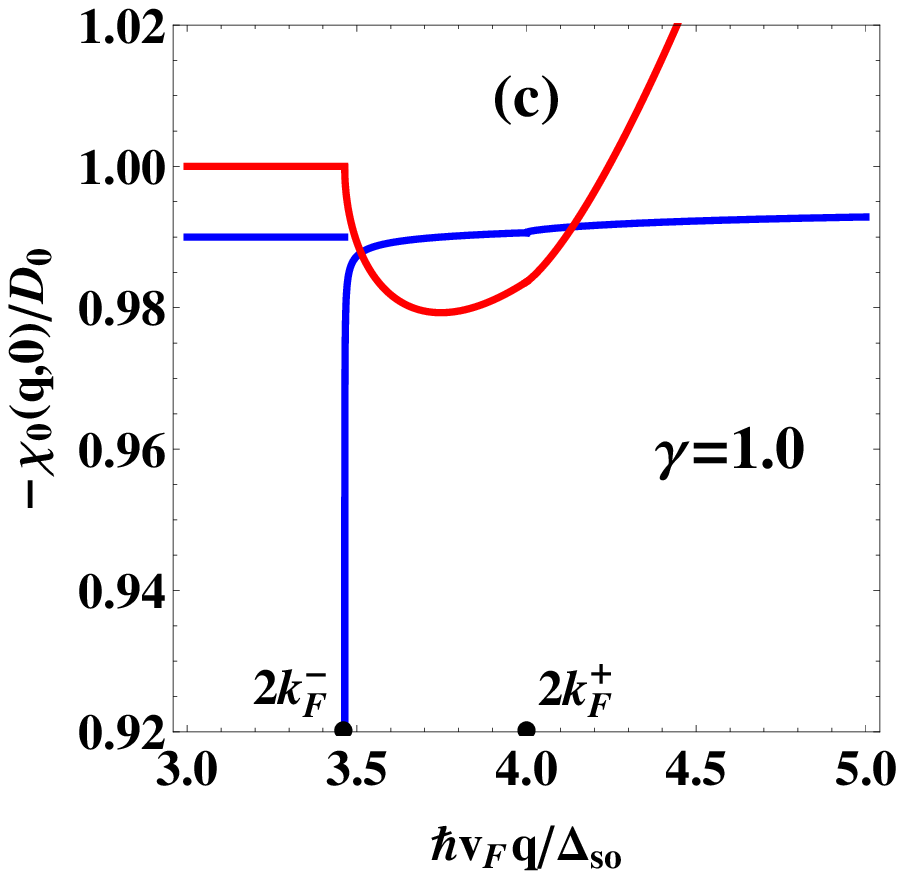}}
\subfigure{\includegraphics[width=4cm]{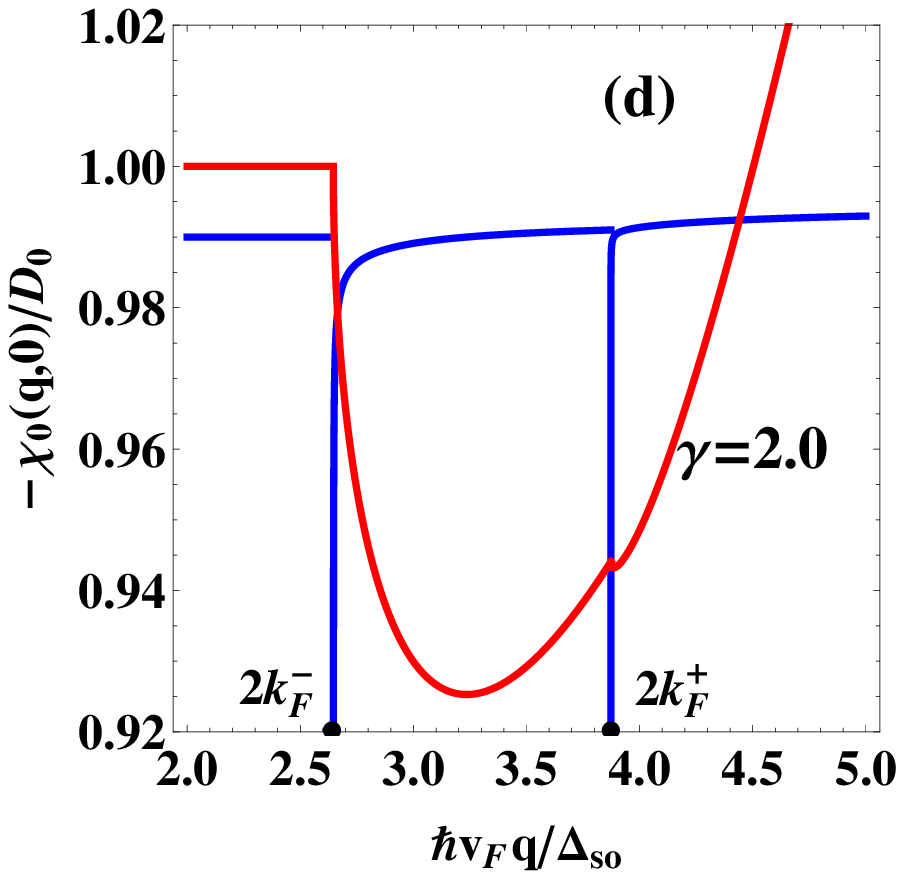}}\\
\caption{(Color online) Static real part of the charge
susceptibility (red line) in units of the density of
states $D_{0}=2\mu/\pi\hbar^2v_{F}^2$, and its corresponding first
derivative is schematically represented by the blue
line.}
\label{Fig2}
\end{figure}

\begin{figure}[htbp]
\centering
\subfigure{\includegraphics[width=4cm]{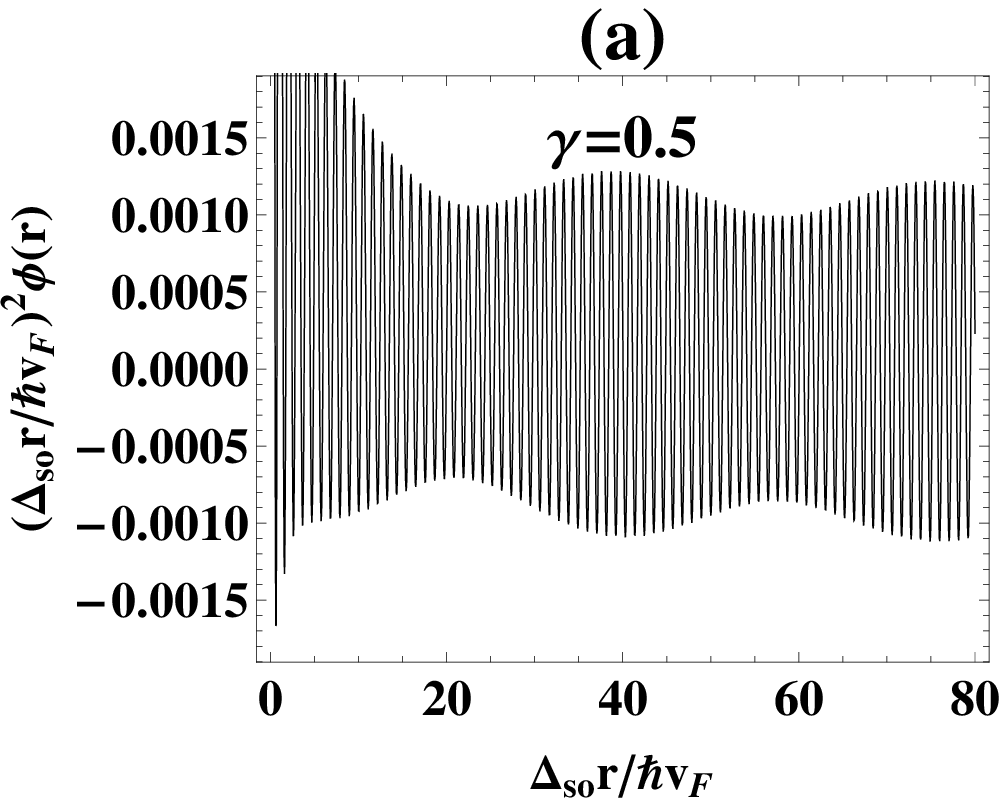}}
\subfigure{\includegraphics[width=4cm]{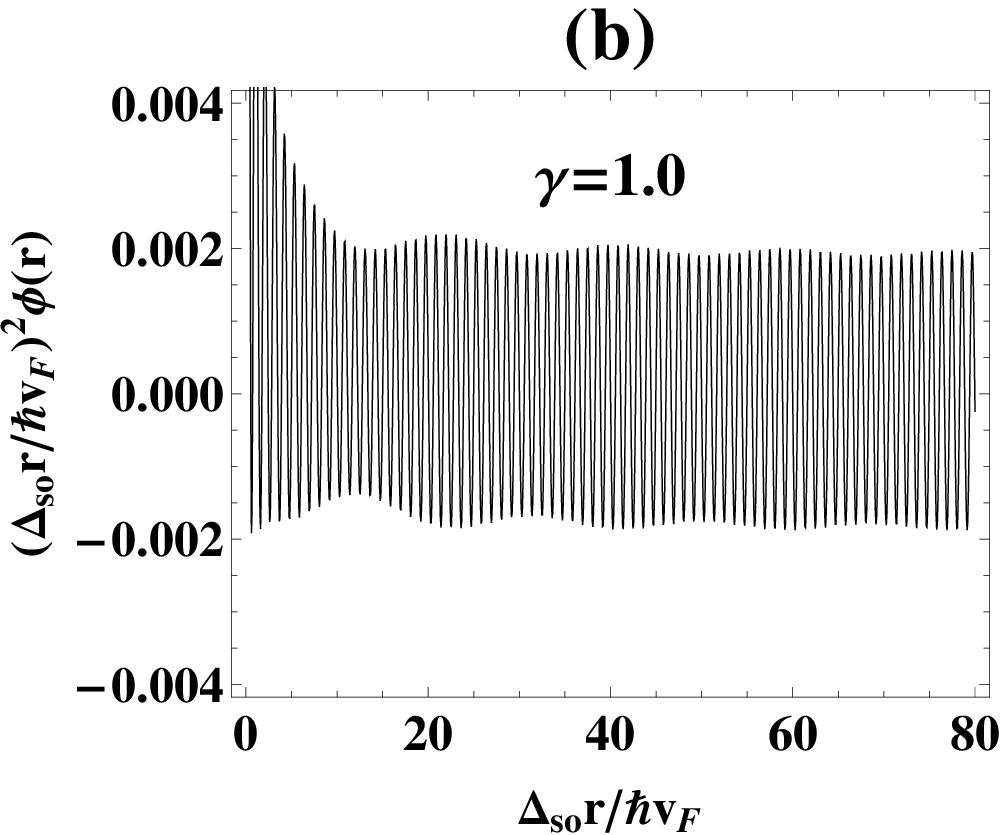}}\\
\subfigure{\includegraphics[width=4cm]{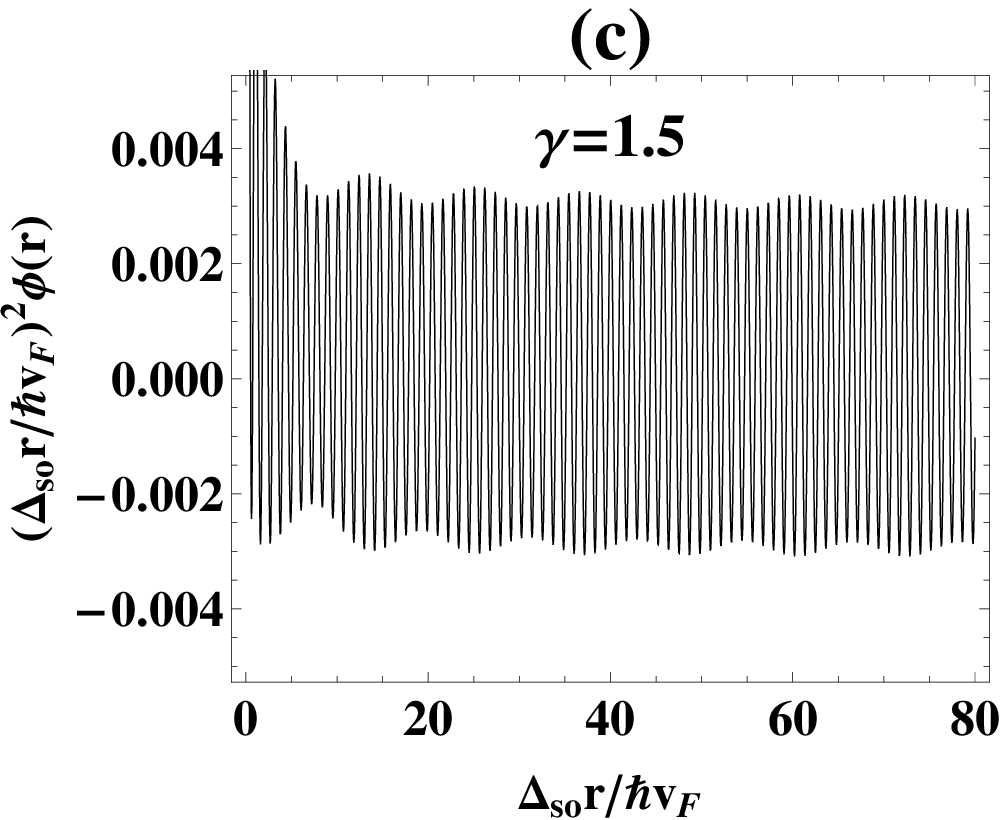}}
\subfigure{\includegraphics[width=4cm]{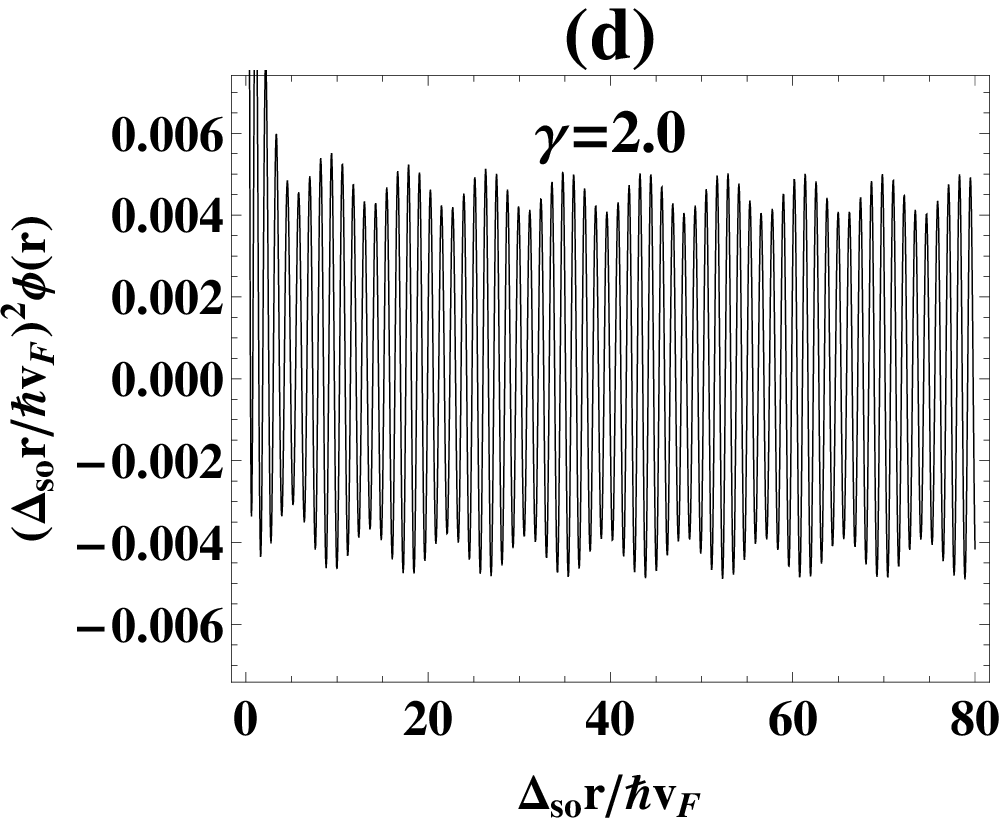}}\\
\caption{The beating behavior of the screened potential
(in units of $Ze\Delta_{\mathrm{so}}/\kappa\hbar v_{F}$)
vs the ratios $\gamma$ for $\kappa=3$,
$\mu=3\Delta_{\mathrm{so}}$, and $v_{F}=c/550$.}
\label{Fig3}
\end{figure}

Therefore, the screened potential can be obtained from Eq.(\ref{phi})
by numerical integration over $q$ (see Fig. \ref{Fig3}).
At a large distance from the charged impurity ($k_{F}^{-}r\gg1$),
there are two main contributions to the screened potential.
The first part is the Thomas-Fermi contribution, $\phi(r)=Ze\hbar^2v_{F}^2/16\kappa\alpha_{\kappa}^{2}\mu^2r^3$
with $\alpha_{\kappa}=e^{2}/\kappa\hbar v_{F}$ the effective
fine structure constant, which is determined by the
long-wavelength ($q\to0$) behavior of the polarization function,
$\chi_{0}(q<2k_{F}^{-},0)=-2\mu/\pi\hbar^2v_{F}^2$. The scale $1/r^{3}$ of the screened
potential at the long-wavelength limit can also be found in the traditional
2D electron gas (2DEG) \cite{Stern}, gapless and gapped graphene
\cite{Wunsch,HwangSarma,Polini,xueFengWang,Pyatkovskiy,ScholzSOC}.

The second part is oscillatory, the Friedel oscillation \cite{Friedel}.
The Lighthill theorem \cite{Lighthill} states that singularities
in the derivatives of the polarization function lead to an algebraic,
and oscillating decay of the screened potential. For the case $\gamma\neq1$
($\Delta_{\pm}\neq0$), the first derivative of the polarization function
is discontinuous at $q=2k_{F}^{\pm}$. For the case $\gamma=1$ ($\Delta_{+}=0$
and $\Delta_{-}\neq0$), the first derivative of the polarization function
is discontinuous at $q=2k_{F}^{-}$ but continuous at $q=2k_{F}^{+}$, and the
second derivative of the polarization function is singular at $q=2k_{F}^{+}$.
Thus, the asymptotic screened potential is a superposition of two kinds of
oscillations, which gives rise to a $beating$ in the oscillatory part of the
screened potential as (see the Supplemental Material \cite{SM}),
\begin{eqnarray}
&&\phi_{\gamma\neq1}(r)\approx
\frac{Z e}{\kappa}\big[F_{-}(r)+F_{+}(r)\big],\label{oscpoten1}\\
&&\phi_{\gamma=1}(r)\approx
\frac{Z e}{\kappa}\big[F_{-}(r)+G_{+}(r)\big],\label{oscpoten2}
\end{eqnarray}
where the functions $F_{\pm}(r)$, $G_{+}(r)$, and $f_{\pm}$
are given by
\begin{eqnarray}
&&F_{\pm}(r)=
-\frac{\alpha_{\kappa}\hbar v_{F}\Delta_{\pm}^{2}}{2\mu\big(\hbar v_{F}k_{F}^{\pm}+2\alpha_{\kappa}\mu f_{\pm}\big)^2}\frac{\sin(2k_{F}^{\pm}r)}{r^2},\\
&&G_{+}(r)=\frac{\alpha_{\kappa}\hbar^2v_{F}^2}{4\mu^2(1+2\alpha_{\kappa}f_{+})^2}
\frac{\cos(2k_{F}^{+}r)}{r^3},\\
&&\hspace{0.5cm}f_{\pm}=1-\frac{1\pm1}{2}\Big(\frac{\sqrt{k_{F}^{+2}-k_{F}^{-2}}}{4k_{F}^{+}}\nonumber\\
&&-\frac{\hbar^2v_{F}^2k_{F}^{+2}-\Delta_{-}^2}{4\hbar v_{F}k_{F}^{+}\mu}
\arctan\frac{\hbar v_{F}\sqrt{k_{F}^{+2}-k_{F}^{-2}}}{\mu}\Big).
\end{eqnarray}
It should be noted that the discrepancy between the oscillatory decay $\sin\left(2k_{F}r\right)/r^{2}$ in gapped graphene \cite{Pyatkovskiy,ScholzSOC}
and $\cos\left(2k_{F}r\right)/r^{3}$ in gapless graphene \cite{Wunsch} is because
of the unique $\pi$ Berry phase of each Dirac point in gapless graphene that
suppresses the backscattering of particles on the Fermi surface during
intravalley scattering \cite{Ando}.

When $\gamma\neq0,\:1$, both the spin-up and spin-down bands are gapped,
and therefore the static polarizability has two cusps and its first
derivative is singular at $q=2k_{F}^{\pm}$ [see Figs. \ref{Fig2}$(\mathrm{b})$
and \ref{Fig2}$(\mathrm{d})$]. The corresponding oscillatory potential is
displayed in Figs. \ref{Fig3}$(\mathrm{a}$, \ref{Fig3}$(\mathrm{c})$, and \ref{Fig3}$(\mathrm{d})$. For a gapped graphene with $\gamma=0$, the two
Fermi wave vectors $k_{F}^{\sigma}$ become degenerate so that only one cusp
of static polarizability remains, and the singularity of its first derivative
locates at $q=2k_F^{0}$ (with $k_F^{+}=k_F^{-}\equiv k_F^{0}$) [see Fig. \ref{Fig2}$(\mathrm{a})$]. Therefore there is only a one-component Friedel
oscillation which does not support beating. Interestingly, at the TPT point
$\gamma=1.0$, the polarization function has a cusp at $q=2k_F^{-}$, and its first derivative is continuous at $q=2k_F^{+}$ but discontinuous at $q=2k_F^{-}$, as
shown in Fig. \ref{Fig2}$(\mathrm{c})$. Thus the screened potential turns out to be a superposition of $\cos\left(2k_{F}^{+}r\right)/r^{3}$
for gapless graphene with $k_{F}^{+}=\mu/\hbar v_F$ and $\sin\left(2k_{F}^{-}r\right)/r^{2}$
for gapped graphene with $k_{F}^{-}=\sqrt{\mu^{2}-\Delta_{\mathrm{so}}^{2}}/\hbar v_F$.
Since the latter overwhelmingly dominates in magnitude at a large distance,
the superposition of these two oscillating parts does not have an indistinguishable beating as shown in Fig. \ref{Fig3}$(\mathrm{b})$. Note that the screened potentials can be captured by our approximative expressions in Eqs.(\ref{oscpoten1}) and
(\ref{oscpoten2}) (more details can be found in the Supplemental Material \cite{SM}).

From Figs. \ref{Fig3}$(\mathrm{a})$-\ref{Fig3}$(\mathrm{d})$, one can clearly see that
as the ratio $\gamma$ approaches $1$, the beating gets gradually faint and disappears
at the TPT point. Once crossing the TPT point,
the beating of the oscillatory potential becomes noticeable. Therefore, the evolution of the beating can be seen as a fingerprint of TPT
in metallic silicene. Note that the beating in the topological phase is usually much longer than the counterpart in the normal phase, which provides us with quantitative evidence of the topological nature of the phase. The Friedel oscillation can be extracted from
the differential tunneling conductance using scanning tunneling
spectroscopy (STS), which has been used to investigate the Friedel
oscillation in a $\sqrt{3}\times\sqrt{3}$ superstructure of silicene
on Ag(111) \cite{STSexp}. Therefore, we expect that further STS measurements
on the Friedel oscillation of silicene could be used to identify the TPT
tuned by the external electrical fields.

\begin{figure}[htbp]
\centering
\subfigure{\includegraphics[width=4cm]{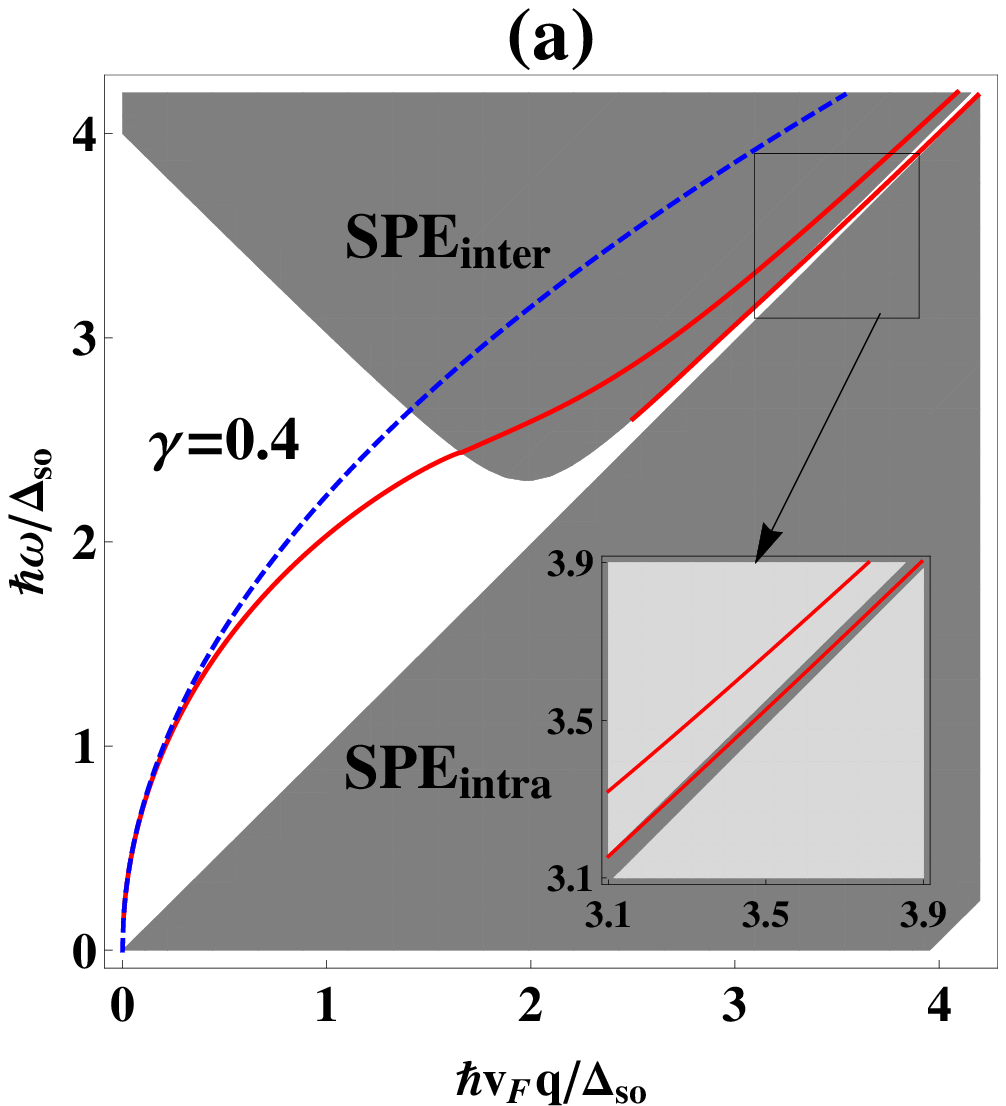}}
\subfigure{\includegraphics[width=4cm]{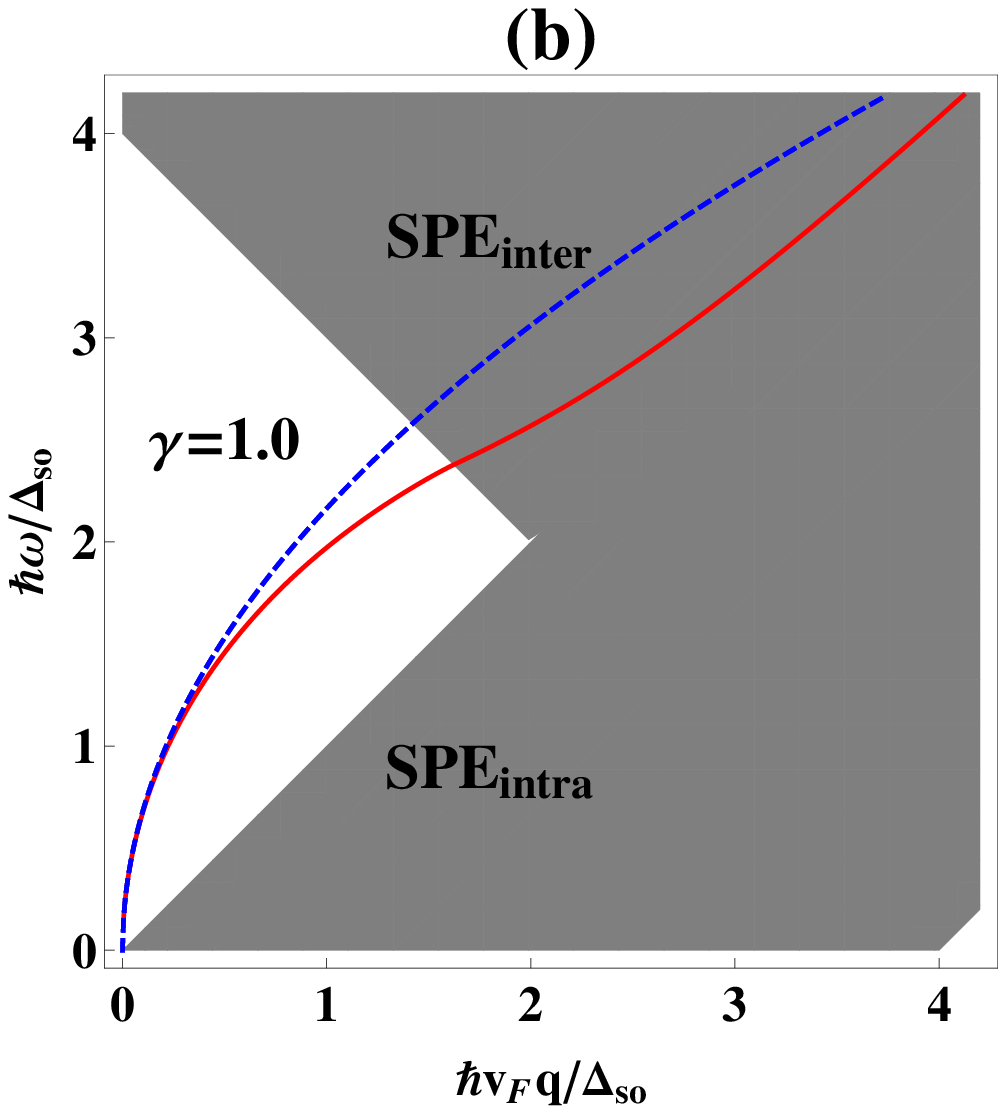}}\\
\subfigure{\includegraphics[width=4cm]{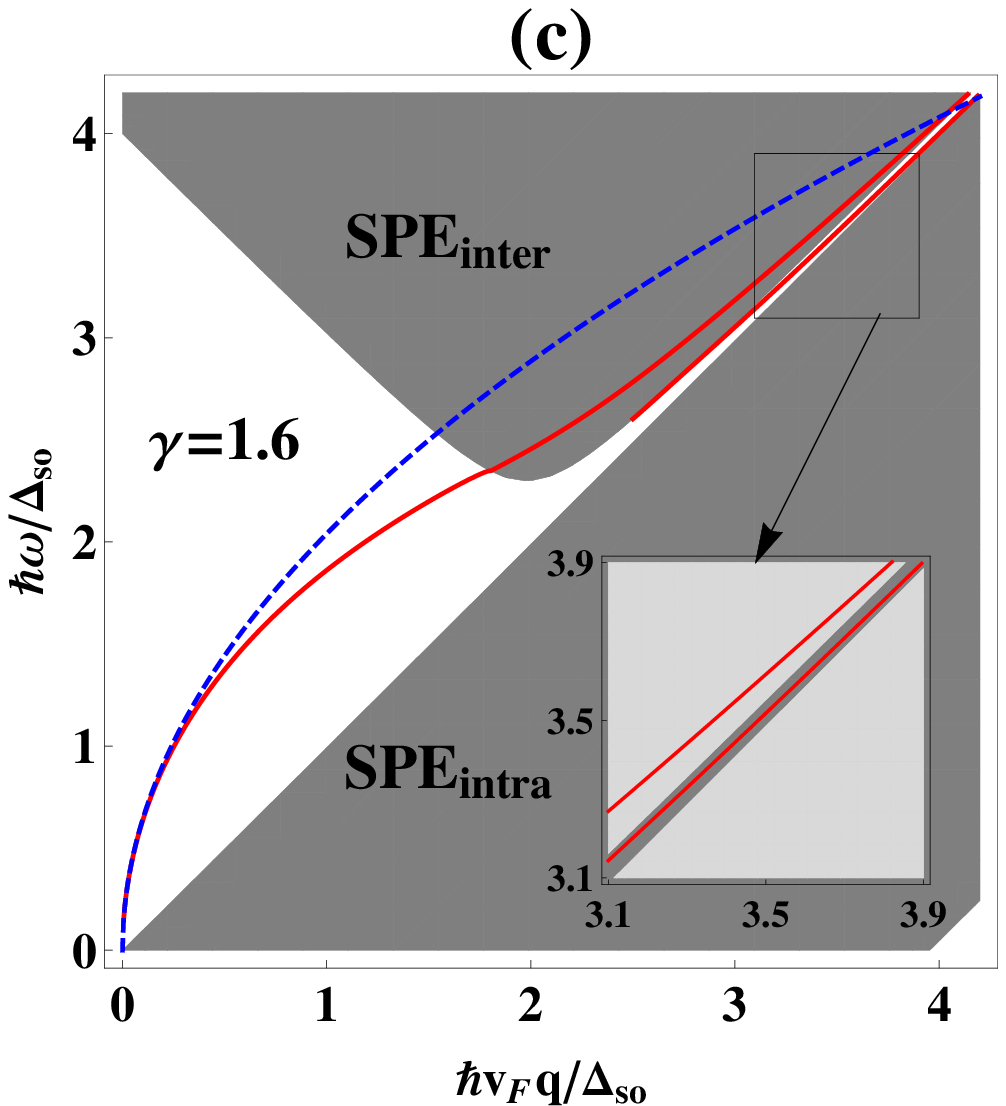}}
\subfigure{\includegraphics[width=4.2cm]{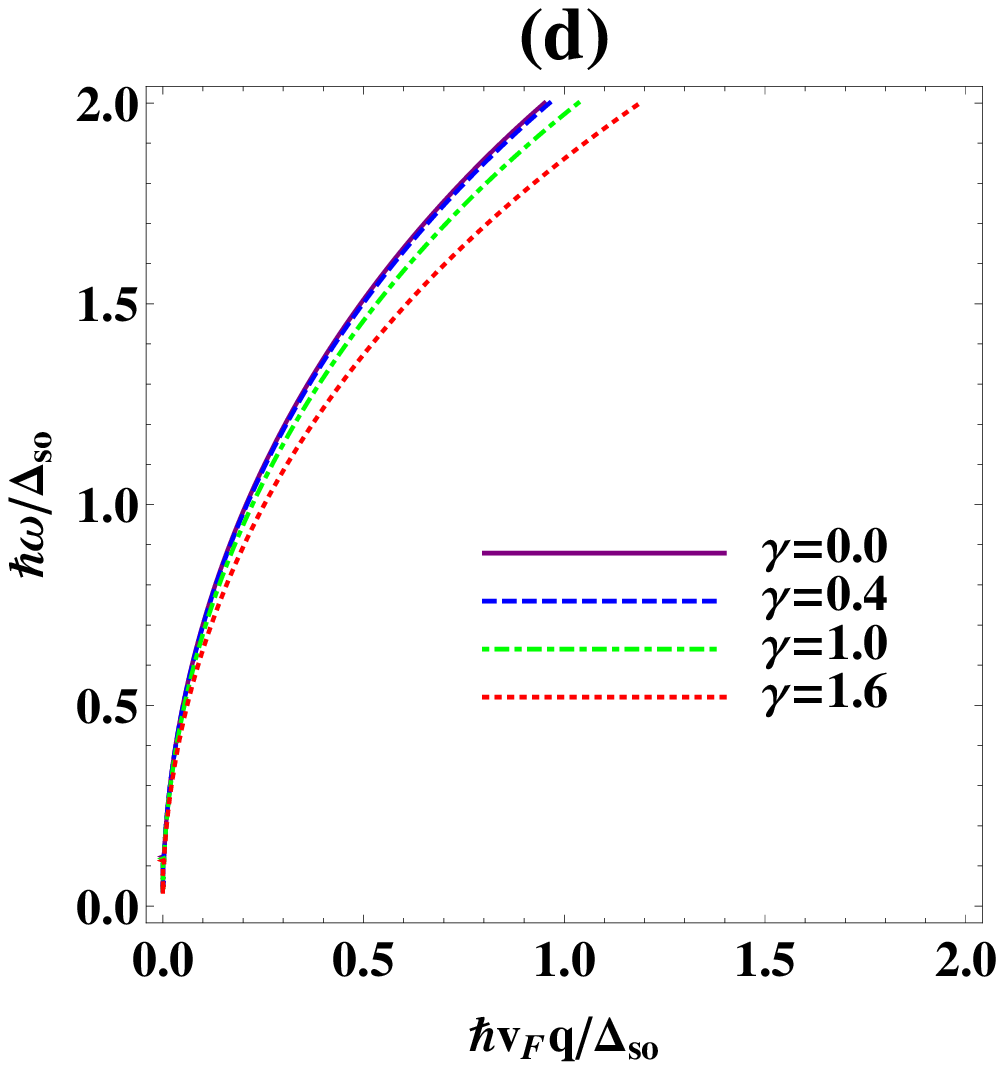}}\\
\caption{(Color online) The numerical exact solution within
RPA (red) and the long-wavelength plasmon frequency (blue),
with the shaded areas representing the interband and the
intraband SPE regions. (d) shows the effect of $\gamma$
on the plasmon frequency. In the inset of (a) or (c), there
is a clear UPM in the SPE gap \cite{NumConst}.}
\label{Fig4}
\end{figure}

We now turn to calculate the plasmon dispersion $\omega_{p}(q)$ by
solving the following equation,
\begin{equation}
\varepsilon\left(q,\omega_{p}-i\delta\right)=0,\label{plasmonexact}
\end{equation}
where $\delta$ is the decay rate of the plasmons. At small energies and momenta $\hbar v_{F}q\ll\hbar\omega\ll\mu$ with $\Delta_{+}\le\Delta_{-}<\mu$,
one can obtain the polarization function in the long-wavelength limit,
\begin{equation}
\chi_{0}\left(q,\omega\right)\approx\mu q^{2}
\left[2-\left(\Delta_{+}^{2}+\Delta_{-}^{2}\right)/\mu^{2}\right]
/(2\pi\hbar^2\omega^{2}).\label{approxexpress}
\end{equation}
 Hence, one immediately obtains the corresponding plasmon frequencies,
\begin{equation}
\omega_{p}^{0}(q)\approx\sqrt{\alpha_{\kappa}\mu v_{F}
\left[2-\left(\Delta_{+}^{2}+\Delta_{-}^{2}\right)/\mu^{2}\right]q/\hbar}.
\label{plasmonlongwave}
\end{equation}
We note that $\omega_{p}^{0}\left(q\right)\propto\sqrt{q}$ is a peculiarity
of 2D plasmon dispersions including both the 2DEG and Dirac fermions.
In Figs. \ref{Fig4}$(\mathrm{a})$-\ref{Fig4}$(\mathrm{c})$, we show the exact numerical plasmon dispersion within RPA (solid red line) and the long-wavelength plasmon (LWP) (dashed blue
line). The SPE region is given by $\mathrm{Im}\left[\chi_{0}\left(q,\omega\right)\right]\neq0$
and displayed as shaded areas in the $(q,\omega)$ space, where the
plasmon is damped into electron-hole pairs (Landau damping). As shown
in Figs. \ref{Fig4}$(\mathrm{a})$-\ref{Fig4}$(\mathrm{c})$, the LWP
frequency $\omega_{p}^{0}$ is well consistent with the exact numerical
solution $\omega_{p}$ in the small wave vector regime.

At the TPT point $\left(\gamma=1\right)$, there exists only one plasmon
mode which is undamped until it enters the interband SPE region. Note
that this branch of the plasmon had been studied extensively in the context
of graphene and a number of exciting applications were proposed for plasmonics
\cite{Grigorenko}. Apart from the TPT point, the exact numerical solution
displays two plasmon modes: One branch of the plasmon is mentioned above,
and the other is a UPM that emerges in the SPE gap [see the insets of Figs. \ref{Fig4}$(\mathrm{a})$-\ref{Fig4}$(\mathrm{c})$].
This UPM can be observed by infrared optical spectroscopy \cite{Basov},
angular-resolved photoemission spectroscopy \cite{Bostwick}, and
other methods \cite{Grigorenko}. It should be noted
that although our UPM has been found in gapped graphene, the observation
in practice is still hard to achieve because of the weak interaction with the substrate or the small intrinsic SOC.
To protect the UPM against the thermal effects and impurity scattering
requires a large SPE gap, which is mainly determined by the SOC gap.
The SOC gaps are about $1.55$ and $24.0$ $\mathrm{meV}$
for silicene and germanene, respectively \cite{PRLYao}. For silicene,
its corresponding energy ranges from $1$ to $10$ $\mathrm{meV}$
by tuning the chemical potential and it can be detected experimentally on
insulating substrates. For germanene,
the energy of UPM is an order of magnitude larger than that of silicene
and is accessed by some experimental methods on both insulating and
metallic substrates \cite{metallicSub}.

Compared with gapped graphene \cite{Pyatkovskiy,xueFengWang,ScholzSOC},
the tunable spin-valley coupled band structure in silicene gives
rise to two additional features in the plasmon spectrum. First, as
shown in Fig. \ref{Fig4}$(\mathrm{d})$, the plasmon frequencies
have redshifted as the ratio $\gamma$ increases by adjusting
the electric field, which can be read out from the long-wavelength
solution in Eq.$\left(\ref{plasmonlongwave}\right)$. Second,
the UPM in the SPE gap decays into a SPE continuum at the TPT point and then
reappears after the system separates from the TPT point. When
closing to the TPT point $\left(\gamma\rightarrow1\right)$, the system
has a very tiny band gap, and the thermal fluctuation or disorder scattering
could smear the small gap and eventually destroy the UPM. Hence, the
disappearance of the UPM can take place in a small region around $\gamma=1$
in the spectral function and help us to identify the TPT point. Therefore,
the fate of the UPM indeed reflects the change of the topology of
energy bands. Furthermore, combining with LWP, the behavior of the plasmons
can clearly characterize the topological nature of the phase. For instance,
when the system is in the normal phase with $1<\gamma$, decreasing the
electric field $E_{z}$ makes the LWP have a violetshift and the UPM disappears
near the TPT point and then reappears in the topological phase. Discussions of
other situations are similar.

In summary, we explored Friedel oscillations and collective excitation
in silicene and germanene within RPA. We emphasize that the tunable
spin-valley coupled band structure could cause some exotic properties
in these two phenomena. We demonstrate that the beating of the screened impurity
potential and the behavior of UPM can both be used as probes for TPT in
silicene and germanene. Our work sheds light on the identification of
TPT via some physical effects. The distinct features of the density oscillations
can also provide us with some hints on how to resolve controversy concerning the
origin of the linear dispersion.

\emph{Note added.} Recently, we became aware of a similar work in this
field \cite{Tabert}.

We are grateful to G. Z. Liu, P. K. Pyatkovskiy, K. H. Wu, D. Xiao,
F. Zhang and F. R. Zhang for favorable discussions and comments. This
work is supported in part by the MOST Project of China under Grant No.
2014CB920903, the Scientific Research Fund of the Education Department
of Sichuan Province under Grant No. 13ZB0157, the NSFC under Grants No.
11074234, No.11274286, No.11174337 and No.11225418, and the U.S. Department of Energy,
Office of Basic Energy Sciences, Materials Sciences and Engineering Division.

\begin{widetext}

\begin{center}
{\textbf{Supplementary material for ``Probing the Topological Phase Transition via Density Oscillations in Silicene and Germanene"}}
\end{center}

\section{preliminary}
In this supplementary material we present the major steps of calculating
the screened potential of silicene at large distantce by following the
method \cite{Harrison}. As indicated in the main text, the screened
potential takes the form
$$\phi(r)=\frac{Z e}{\kappa}\int_0^\infty
\frac{J_0(q r)}{\varepsilon(q,0)}d q,\eqno(\mathrm{S.1})$$
where the static dielectric function $\varepsilon(q,0)=1-V(q)\chi_{0}(q,0)$
with $V(q)=2\pi e^2/\kappa q$. When the Fermi level cuts two Fermi
surfaces ($\Delta_{+}\le\Delta_{-}<\mu$), according to Eq.($6$) in
the main text the static dielectric function is of the form
$$\varepsilon(q,0)=1+\frac{4\alpha_{\kappa}\mu}{\hbar v_{F}q}
\Big[1-\sum_{\sigma=\pm}\theta(q-2k_{F}^{\sigma})
\Big(\frac{\sqrt{q^{2}-4k_{F}^{\sigma2}}}{4q}
-\frac{\hbar^2v_{F}^2q^{2}-4\Delta_{\sigma}^{2}}{8\hbar v_{F}q\mu}
\arctan\frac{\hbar v_{F}
\sqrt{q^{2}-4k_{F}^{\sigma2}}}{2\mu}\Big)\Big].\eqno(\mathrm{S.2})$$
For the sake of simplicity, we introduce some dimensionless quantities,
$$\tilde{q}=\frac{\hbar v_F q}{\Delta_{so}},
\hspace{1cm}\tilde{r}=\frac{\Delta_{so} r}{\hbar v_F},
\hspace{1cm}\tilde{\mu}=\frac{\mu}{\Delta_{so}},
\hspace{1cm}\tilde{\Delta}_{\sigma}=\frac{\Delta_{\sigma}}{\Delta_{so}},
\hspace{1cm}\tilde{k}_F^{\sigma}=\frac{\hbar v_F k_F^{\sigma}}{\Delta_{so}}.
\eqno(\mathrm{S.3})$$
Note that $\tilde{\Delta}_{\pm}=|\gamma\mp1|/2$ and $\tilde{k}_{F}^{\pm}=\sqrt{\tilde{\mu}^{2}-\tilde{\Delta}_{\pm}^{2}}$.
In terms of $\tilde{\mu}$, $\tilde{\Delta}_{\sigma}$, $\tilde{k}_F^{\sigma}$, $\tilde{q}$ and $\tilde{r}$, the screened potential is of the form
$$\phi(r)=\frac{Z e \Delta_{so}}{\kappa\hbar v_F}\psi(\tilde{r}),\eqno(\mathrm{S.4})$$
where
$$\psi(\tilde{r})=\int_0^\infty\frac{J_0(\tilde{q}\tilde{r})}
{\tilde{\varepsilon}(\tilde{q},0)}d\tilde{q},\eqno(\mathrm{S.5})$$
and the static dielectric function can be expressed as,
$$\tilde{\varepsilon}(\tilde{q},0)=1+\frac{4\alpha_{\kappa}\tilde{\mu}}{\tilde{q}}
\Big[1-\sum_{\sigma=\pm}\theta(\tilde{q}-2\tilde{k}_F^{\sigma})
\Big(\frac{\sqrt{\tilde{q}^{2}-4\tilde{k}_F^{\sigma2}}}{4\tilde{q}}
-\frac{\tilde{q}^{2}-4\tilde{\Delta}_{\sigma}^{2}}{8\tilde{q}\tilde{\mu}}
\arctan\frac{\sqrt{\tilde{q}^{2}-4\tilde{k}_F^{\sigma2}}}{2\tilde{\mu}}\Big)\Big].
\eqno(\mathrm{S.6})$$
It can be checked that (keep in mind $\tilde{k}_{F}^{-}\le \tilde{k}_{F}^{+}$)
$$\tilde{\varepsilon}(\tilde{q}<2\tilde{k}_{F}^{-},0)=
1+4\alpha_{\kappa}\tilde{\mu}/\tilde{q},\eqno(\mathrm{S.7})$$
$$\tilde{\varepsilon}(\tilde{q}\to\infty,0)=1+\alpha_{\kappa}\pi/2.\eqno(\mathrm{S.8})$$
For convenience, we introduce two auxiliary functions which will be useful later. The first is
$$\bar{J}_0(z)\equiv
z ~_1F_2\big(\frac{1}{2};1,\frac{3}{2};-\frac{1}{4} z^2\big)-1,
\eqno(\mathrm{S.9})$$
which satisfies $\frac{d\bar{J}_0(z)}{d z}=J_0(z)$, where $_mF_n(\alpha_1,\cdots,\alpha_m;\beta_1,\cdots,\beta_n;z)$
is the generalized hypergeometric function \cite{Gradshteyn}.
For $z\gg1$ the asymptotic expression reads \cite{Miller},
$$\bar{J}_0(z)\approx\sqrt{\frac{2}{\pi z}}\sin(z-\frac{\pi}{4}).\eqno(\mathrm{S.10})$$
The second is
$$\hat{J}_0(z)=\frac{1}{2}z^2~_1F_2\big(\frac{1}{2};\frac{3}{2},2;-\frac{1}{4} z^2\big)-z,\eqno(\mathrm{S.11})$$
which satisfies $\frac{d\hat{J}_0(z)}{d z}=\bar{J}_0(z)$.
For $z\gg1$ the asymptotic expression reads \cite{Miller},
$$\hat{J}_0(z)\approx\sqrt{\frac{2}{\pi z}}\cos(z+\frac{3}{4}\pi).\eqno(\mathrm{S.12})$$

\section{The evaluation of the oscillatory part of screened potential}
In this section, we calculate the oscillatory part of the screened potential
at large distance from a charged impurity ($\tilde{k}_{F}^{-}r\gg1$).
Integrating by parts leads to
$$\psi(\tilde{r})=\frac{1}{\tilde{r}}\int_0^\infty\frac{d \bar{J}_0(\tilde{q}\tilde{r})}{\tilde{\varepsilon}(\tilde{q},0)}
=\underline{\frac{1}{\tilde{r}}\left[\frac{\bar{J}_0(\tilde{q}\tilde{r})}
{\tilde{\varepsilon}(\tilde{q},0)}\right]_0^\infty}
+\frac{1}{\tilde{r}}\int_0^\infty\frac{\bar{J}_0(\tilde{q}\tilde{r})}
{\tilde{\varepsilon}(\tilde{q},0)^2}\frac{d \tilde{\varepsilon}(\tilde{q},0)}{d \tilde{q}}d \tilde{q},\eqno(\mathrm{S.13})$$
From Eqs.($\mathrm{S.7}$), ($\mathrm{S.8}$) and noticing that $\bar{J}_0(z\to0)=-1$
and $\bar{J}_0(z\to\infty)=0$ we find the underline term vanishes, then
$$\psi(\tilde{r})=\frac{1}{\tilde{r}}\int_0^\infty\frac{\bar{J}_0(\tilde{q}\tilde{r})}
{\tilde{\varepsilon}(\tilde{q},0)^2}\frac{d \tilde{\varepsilon}(\tilde{q},0)}{d \tilde{q}}d \tilde{q}.\eqno(\mathrm{S.14})$$
According to Lighthill theorem \cite{Lighthill}, the contribution
of the integral comes only from the non-analyticity of the polarization
function at $\tilde{q}=2\tilde{k}_{F}^{\pm}$ where its derivatives
are discontinuous. In this case, $\tilde{q}=2\tilde{k}_{F}^{-}$
and $\tilde{q}=2\tilde{k}_{F}^{+}$ are two singularities of the
derivatives of the static polarization function (when $\gamma=0$
the two singularities degenerate), therefore one can make the
decomposition to pick up the main contribution
$$\psi(\tilde{r})
\approx\psi_{\tilde{k}_{F}^{-}}(\tilde{r})+\psi_{\tilde{k}_{F}^{+}}(\tilde{r}).
\eqno(\mathrm{S.14})$$
Around $\tilde{q}=2\tilde{k}_{F}^{\pm}$ the static dielectric function reads
$$\tilde{\varepsilon}(\tilde{q}\approx2\tilde{k}_F^{\pm},0)\approx
1+\frac{2\alpha_{\kappa}\tilde{\mu}}{\tilde{k}_F^{\pm}}\tilde{f}_{\pm},
\eqno(\mathrm{S.15})$$
which is finite with
$$\tilde{f}_{\pm}=1-\frac{1\pm1}{2}
\left(\frac{\sqrt{\tilde{k}_F^{+2}-\tilde{k}_F^{-2}}}{4\tilde{k}_F^{+}}
-\frac{\tilde{k}_F^{+2}-\tilde{\Delta}_{-}^2}{4\tilde{k}_F^{+}\tilde{\mu}}
\arctan\frac{\sqrt{\tilde{k}_F^{+2}-\tilde{k}_F^{-2}}}{\tilde{\mu}}\right).
\eqno(\mathrm{S.16})$$

\subsection{For the case with $\gamma\neq1$}

For $\gamma\neq1$ (namely, $\tilde{\Delta}_{\pm}\neq0$),
the first derivative of the static dielectric function
$\frac{d \tilde{\varepsilon}(\tilde{q},0)}{d \tilde{q}}$ takes a form $\frac{\theta(\tilde{q}-2\tilde{k}_F^{\pm})}{\sqrt{\tilde{q}-2\tilde{k}_F^{\pm}}}$
around $\tilde{q}=2\tilde{k}_{F}^{\pm}$, and hence is divergent at $\tilde{q}=2\tilde{k}_{F}^{\pm}$. In addition, the integrand $\frac{\bar{J}_0(\tilde{q}\tilde{r})}{\tilde{\varepsilon}(\tilde{q},0)^2}
\frac{d \tilde{\varepsilon}(\tilde{q},0)}{d \tilde{q}}$ in Eq.($\mathrm{S.14}$)
and its first derivative behave well at both $\tilde{q}=0$ and $\tilde{q}=\infty$.
Therefore, according to Lighthill theorem,
$$\psi_{\tilde{k}_{F}^{\pm}}(\tilde{r})
\approx\frac{1}{\tilde{r}}\int_0^\infty\frac{\bar{J}_0(\tilde{q}\tilde{r})}
{\tilde{\varepsilon}(\tilde{q},0)^2}\frac{d \tilde{\varepsilon}(\tilde{q},0)}{d \tilde{q}}d \tilde{q}\Big|_{\tilde{q}\approx2\tilde{k}_{F}^{\pm}},\eqno(\mathrm{S.17})$$
with
$$\frac{d \tilde{\varepsilon}(\tilde{q},0)}{d \tilde{q}}\Big|_{\tilde{q}\approx2\tilde{k}_{F}^{\pm}}
\approx-\frac{\alpha_{\kappa}\tilde{\Delta}_{\pm}^{2}}{2\tilde{\mu}\tilde{k}_{F}^{\pm}\sqrt{\tilde{k}_{F}^{\pm}}}
\frac{\theta(\tilde{q}-2\tilde{k}_{F}^{\pm})}{\sqrt{\tilde{q}-2\tilde{k}_{F}^{\pm}}}.
\eqno(\mathrm{S.18})$$

Substituting Eqs.($\mathrm{S.10}$), ($\mathrm{S.15}$) and ($\mathrm{S.18}$) into Eq.($\mathrm{S.17}$), and taking into account the integration
$$\int_0^\infty
\frac{\sin(\tilde{q}\tilde{r}-\frac{\pi}{4})
\theta(\tilde{q}-2\tilde{k}_{F}^{\pm})}
{\sqrt{\tilde{q}-2\tilde{k}_{F}^{\pm}}}d \tilde{q}
=\sqrt{\frac{\pi}{\tilde{r}}}\sin(2\tilde{k}_{F}^{\pm}\tilde{r}),\eqno(\mathrm{S.19})$$
we obtain the oscillatory part of the gapped case
$$\psi_{\tilde{k}_{F}^{\pm}}(\tilde{r})
\approx-\frac{\alpha_{\kappa}\tilde{\Delta}_{\pm}^{2}}{2\tilde{\mu}
(\tilde{k}_{F}^{\pm}+2\alpha_{\kappa}\tilde{\mu}\tilde{f}_{\pm})^2}
\frac{\sin(2\tilde{k}_{F}^{\pm}\tilde{r})}{\tilde{r}^2}.\eqno(\mathrm{S.20})$$

\subsection{For the case with $\gamma=1$}
For $\gamma=1$ (namely, $\tilde{\Delta}_{-}=1$ and $\tilde{\Delta}_{+}=0$), at $\tilde{q}=2\tilde{k}_{F}^{-}$ the first derivative of the static polarization
function is divergent while at $\tilde{q}=2\tilde{k}_{F}^{+}=2\tilde{\mu}$ the
first derivative of the static polarization function is continuous but the
second derivative is divergent. Consequently we are able to consider the first derivative around $\tilde{q} =2\tilde{k}_{F}^{-}$ only but must take the second derivative around $\tilde{q}=2\tilde{k}_{F}^{+}=2\tilde{\mu}$ into account.

For $\tilde{q}\approx2\tilde{k}_{F}^{-}$, from Eq.($\mathrm{S.20}$) we have
$$\psi_{\tilde{k}_{F}^{-}}(\tilde{r})
\approx-\frac{\alpha_{\kappa}\tilde{\Delta}_{-}^{2}}{2\tilde{\mu}
(\tilde{k}_{F}^{-}+2\alpha_{\kappa}\tilde{\mu}\tilde{f}_{-})^2}
\frac{\sin(2\tilde{k}_{F}^{-}\tilde{r})}{\tilde{r}^2}.\eqno(\mathrm{S.21})$$

For $\tilde{q}\approx2\tilde{k}_{F}^{+}$ we consider the second derivative
$$\psi_{\tilde{k}_{F}^{+}}(\tilde{r})
\approx\frac{1}{\tilde{r}}\int_0^\infty\frac{\bar{J}_0(\tilde{q}\tilde{r})}
{\tilde{\varepsilon}(\tilde{q},0)^2}\frac{d \tilde{\varepsilon}(\tilde{q},0)}{d \tilde{q}}d \tilde{q}\Big|_{\tilde{q}\approx2\tilde{k}_{F}^{+}}
=\frac{1}{\tilde{r}^2}\int_0^\infty G(\tilde{q})d\hat{J}_0(\tilde{q}\tilde{r})\Big|_{\tilde{q}\approx2\tilde{k}_{F}^{+}}$$
$$=\left\{\underline{\frac{1}{\tilde{r}^2} \left[G(\tilde{q})\hat{J}_0(\tilde{q}\tilde{r})\right]_0^\infty}
-\frac{1}{\tilde{r}^2}\int_0^\infty\hat{J}_0(\tilde{q}\tilde{r})\frac{d G(\tilde{q})}{d \tilde{q}}d\tilde{q}\right\}\Big|_{\tilde{q}\approx2\tilde{k}_{F}^{+}},\eqno(\mathrm{S.22})$$
where
$$G(\tilde{q})=\frac{1}
{\tilde{\varepsilon}(\tilde{q},0)^2}\frac{d \tilde{\varepsilon}(\tilde{q},0)}{d \tilde{q}}.\eqno(\mathrm{S.23})$$
It can be checked that the underlined term vanishes from $\hat{J}_0(z\to0)=\hat{J}_0(z\to\infty)=0$ and Eqs.($\mathrm{S.7}$), ($\mathrm{S.8}$), then
$$\psi_{\tilde{k}_{F}^{+}}(\tilde{r})
=-\frac{1}{\tilde{r}^2}\int_0^\infty\hat{J}_0(\tilde{q}\tilde{r})\frac{d G(\tilde{q})}{d \tilde{q}}d\tilde{q}\Big|_{\tilde{q}\approx2\tilde{k}_{F}^{+}}.\eqno(\mathrm{S.24})$$
It is noted that the integrand $\hat{J}_0(\tilde{q}\tilde{r})
\frac{d G(\tilde{q})}{d \tilde{q}}$ in Eq.($\mathrm{S.24}$)
and its first derivative behave well at $\tilde{q}=0$ and $\tilde{q}=\infty$.
Because $\frac{d \tilde{\varepsilon}(\tilde{q},0)}{d \tilde{q}}
\Big|_{\tilde{q}\approx2\tilde{k}_{F}^{+}}$ is finite, we obtain
$$\frac{d G(\tilde{q})}{d \tilde{q}}\Big|_{\tilde{q}\approx2\tilde{k}_{F}^{+}}\approx
\frac{1}{\tilde{\varepsilon}(\tilde{q},0)^{2}}\frac{d^2 \tilde{\varepsilon}(\tilde{q},0)}{d \tilde{q}^2}\Big|_{\tilde{q}\approx2\tilde{k}_{F}^{+}}.\eqno(\mathrm{S.25})$$
with
$$\frac{d^2 \tilde{\varepsilon}(\tilde{q},0)}{d \tilde{q}^2}
\Big|_{\tilde{q}\approx2\tilde{k}_{F}^{+}=2\mu}
\approx\frac{\alpha_{\kappa}}{4\tilde{\mu}\sqrt{\tilde{\mu}}}
\frac{\theta(\tilde{q}-2\tilde{\mu})}{\sqrt{\tilde{q}-2\tilde{\mu}}}.\eqno(\mathrm{S.26})$$
Therefore substituting Eqs.($\mathrm{S.12}$), ($\mathrm{S.25}$) and ($\mathrm{S.26}$) into Eq.($\mathrm{S.24}$), and using the integration
$$\int_0^\infty
\frac{\cos(\tilde{q}\tilde{r}+\frac{3}{4}\pi)
\theta(\tilde{q}-2\tilde{k}_{F}^{+})}
{\sqrt{\tilde{q}-2\tilde{k}_{F}^{+}}}d \tilde{q}
=-\sqrt{\frac{\pi}{\tilde{r}}}\cos(2\tilde{k}_{F}^{+}\tilde{r}),\eqno(\mathrm{S.27})$$
we obtain that
$$\psi_{\tilde{k}_{F}^{+}}(\tilde{r})
\approx\frac{\alpha_{\kappa}}{4\tilde{\mu}^2(1+2\alpha_{\kappa}\tilde{f}_{+})^{2}}
\frac{\cos(2\tilde{k}_{F}^{+}\tilde{r})}{\tilde{r}^3}.\eqno(\mathrm{S.28})$$

\section{summary}
For $\gamma\neq1$, the screened potential is
$$\phi_{\gamma\neq1}(r)\approx
\frac{Z e \Delta_{so}}{\kappa\hbar v_F}\sum_{\sigma=\pm}
-\frac{\alpha_{\kappa}\tilde{\Delta}_{\sigma}^{2}}{2\tilde{\mu}
(\tilde{k}_{F}^{\sigma}+2\alpha_{\kappa}\tilde{\mu}\tilde{f}_{\sigma})^2}
\frac{\sin(2\tilde{k}_{F}^{\sigma}\tilde{r})}{\tilde{r}^2}.\eqno(\mathrm{S.29})$$
\begin{figure}[htbp]
\centering
\subfigure{\includegraphics[width=4cm]{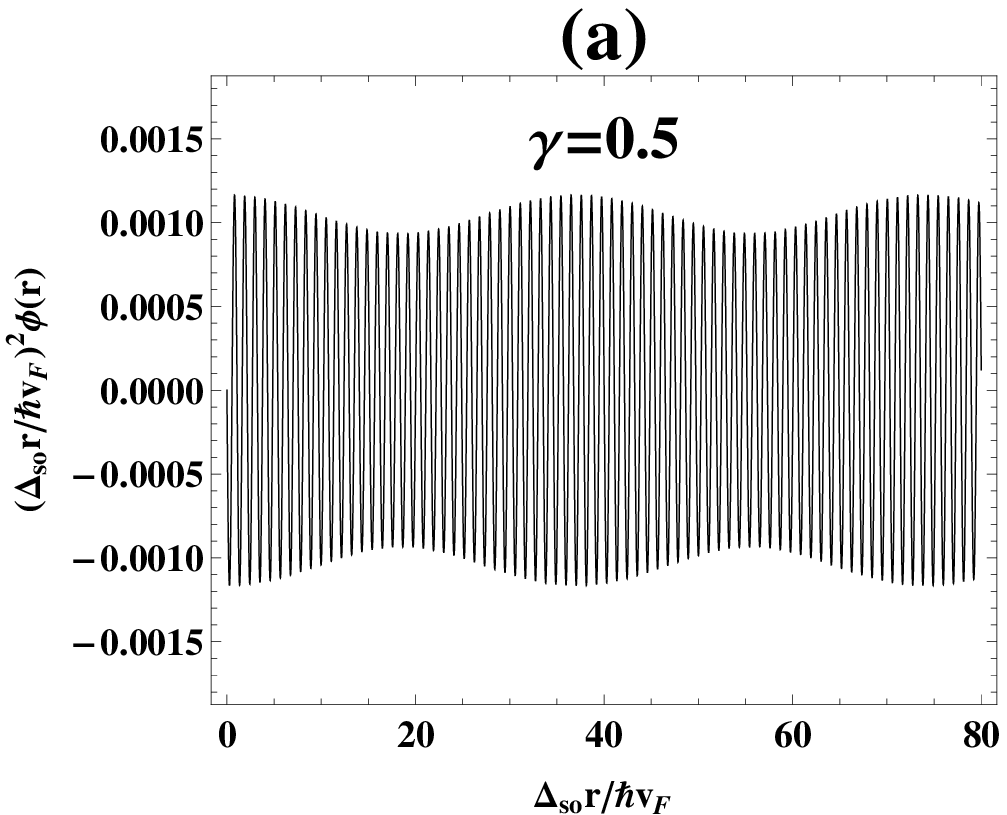}}
\subfigure{\includegraphics[width=4cm]{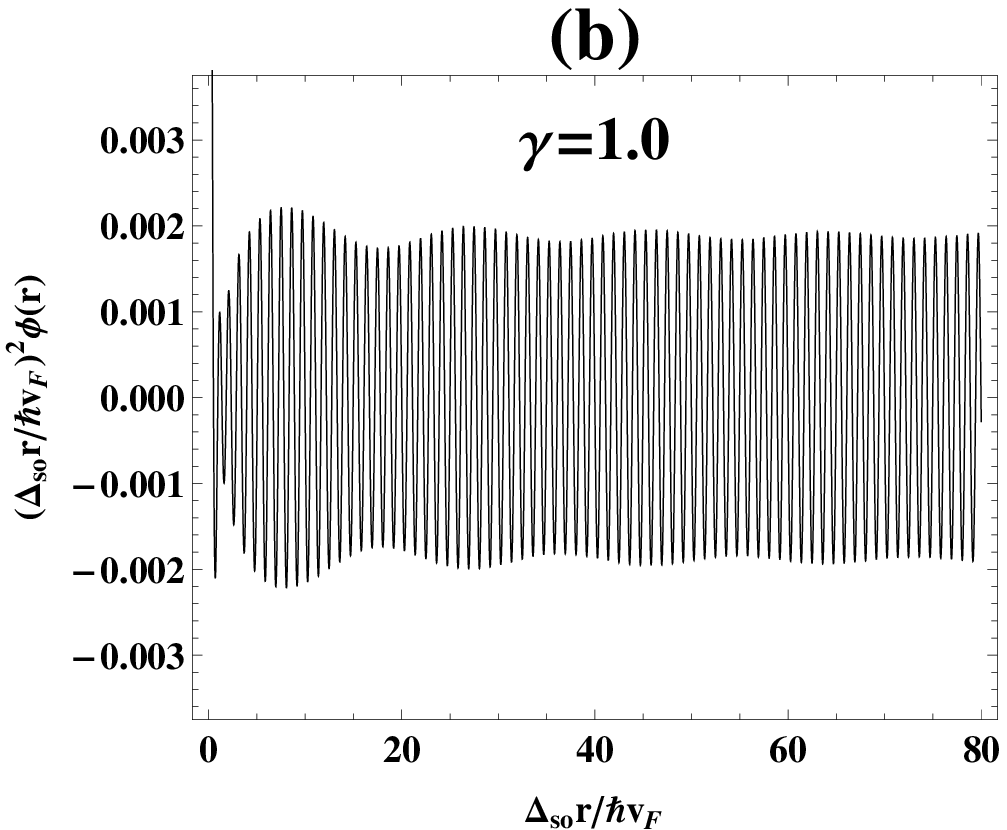}}\\
\subfigure{\includegraphics[width=4cm]{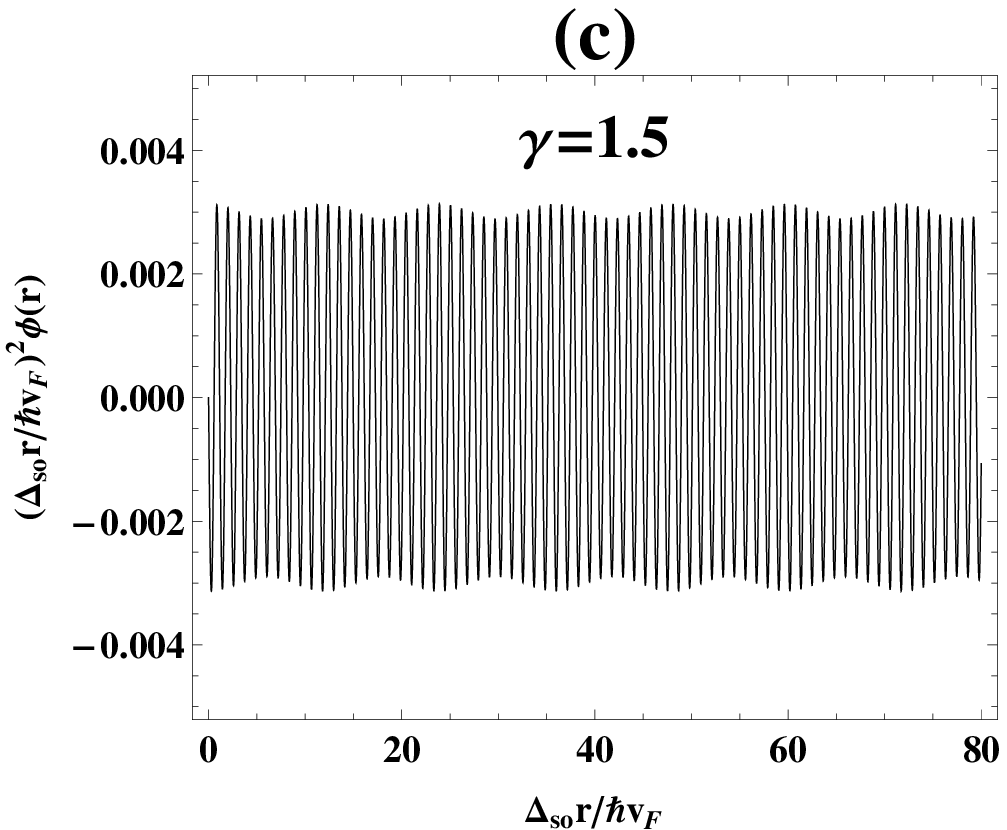}}
\subfigure{\includegraphics[width=4cm]{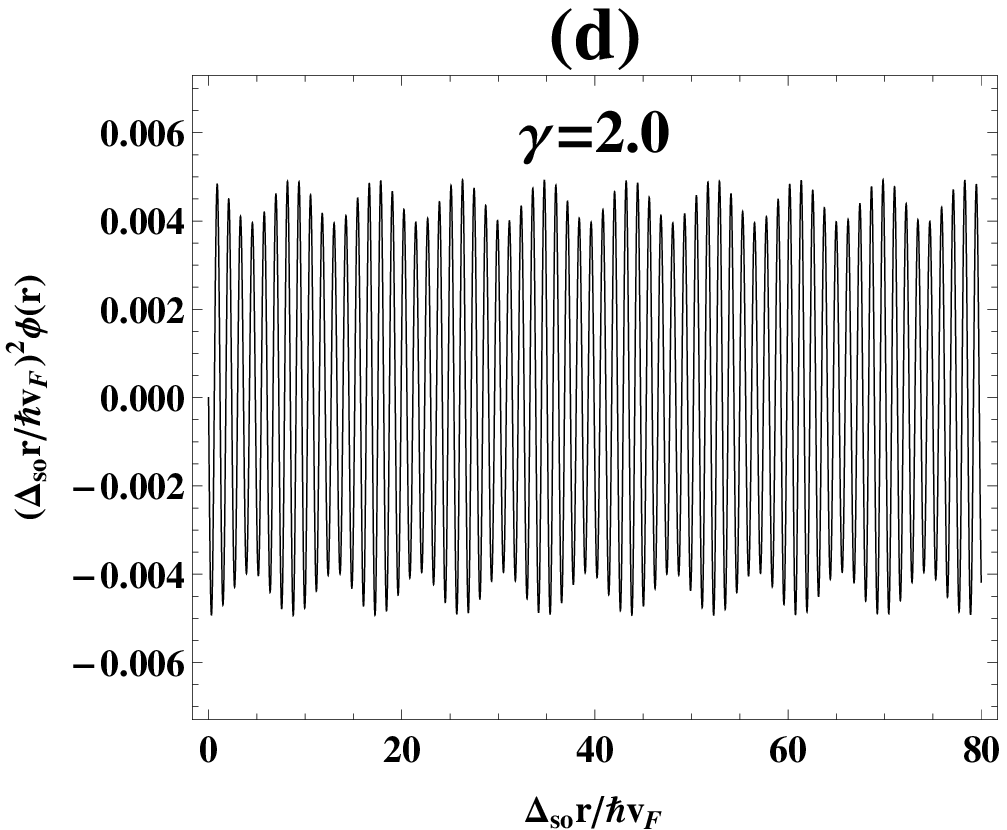}}\\
\caption{The beating behavior of asymptotic screened potential
(in units of $Ze\Delta_{so}/\kappa\hbar v_{F}$) versus the ratios
$\gamma$ by using the Lighthill theorem. We set $\kappa=3$,
$\mu=3\Delta_{\mathrm{so}}$ and $v_{F}=c/550$.}
\label{FigS1}
\end{figure}
For $\gamma=1$, $\tilde{\Delta}_{-}=1$, $\tilde{\Delta}_{+}=0$ ($\tilde{k}_{F}^{-}=\sqrt{\tilde{\mu}^2-1}$, $\tilde{k}_{F}^{+}=\tilde{\mu}$),
the screened potential reads
$$\phi_{\gamma=1}(r)=\frac{Z e \Delta_{so}}{\kappa\hbar v_F}\Big(
-\frac{\alpha_{\kappa}\tilde{\Delta}_{-}^{2}}{2\tilde{\mu}
(\tilde{k}_{F}^{-}+2\alpha_{\kappa}\tilde{\mu}\tilde{f}_{-})^2}
\frac{\sin(2\tilde{k}_{F}^{-}\tilde{r})}{\tilde{r}^2}
+\frac{\alpha_{\kappa}}{4\tilde{\mu}^2(1+2\alpha_{\kappa}\tilde{f}_{+})^{2}}
\frac{\cos(2\tilde{\mu}\tilde{r})}{\tilde{r}^3}\Big).\eqno(\mathrm{S.30})$$
In terms of $\mu$, $\Delta_{\sigma}$, $k_{F}^{\sigma}$ and $r$,
the screened potential reads
$$\phi_{\gamma\neq1}(r)\approx
\frac{Z e}{\kappa}\big[F_{-}(r)+F_{+}(r)\big],\eqno(\mathrm{S.31})$$
$$\phi_{\gamma=1}(r)\approx
\frac{Z e}{\kappa}\big[F_{-}(r)+G_{+}(r)\big],\eqno(\mathrm{S.32})$$
where the functions $F_{\pm}(r)$, $G_{+}(r)$ and $f_{\pm}$ are given by
$$F_{\pm}(r)=-\frac{\alpha_{\kappa}\hbar v_{F}\Delta_{\pm}^{2}}{2\mu\big(\hbar v_{F}k_{F}^{\pm}+2\alpha_{\kappa}\mu f_{\pm}\big)^2}\frac{\sin(2k_{F}^{\sigma}r)}{r^2},\eqno(\mathrm{S.33})$$
$$G_{+}(r)=\frac{\alpha_{\kappa}\hbar^2v_{F}^2}{4\mu^2(1+2\alpha_{\kappa}f_{+})^2}
\frac{\cos(2k_{F}^{+}r)}{r^3},\eqno(\mathrm{S.34})$$
$$f_{\pm}=1-\frac{1\pm1}{2}\Big(\frac{\sqrt{k_{F}^{+2}-k_{F}^{-2}}}{4k_{F}^{+}}
-\frac{\hbar^2v_{F}^2k_{F}^{+2}-\Delta_{-}^2}{4\hbar v_{F}k_{F}^{+}\mu}
\arctan\frac{\hbar v_{F}\sqrt{k_{F}^{+2}-k_{F}^{-2}}}{\mu}\Big).\eqno(\mathrm{S.35})$$

At last, we show that the analytical result in Eqs.($\mathrm{S.31}$) and ($\mathrm{S.32}$)(see Fig. \ref{FigS1} in this supplementary material) is
in good agreement with that in Fig. \ref{Fig3} in the main text in the large $\Delta_{so}r/\hbar v_{F}$ region.

\end{widetext}


\begin{thebibliography}{References}
\bibitem{Si2DGe} Because of the similarity between the germanene
and the silicene, for the remainder of this Rapid Communication,
we will not mention the germanene unless it is necessary.
\bibitem{RelDieConst}  The effective dielectric constant $\kappa=\left(\kappa_{a}+\kappa_{b}\right)/2$
is the average screening contribution of any dielectrics above $\left(\kappa_{a}\right)$
and below $\left(\kappa_{b}\right)$ the sample. The different
values of $\kappa$ do not modify the qualitative behavior.
\bibitem{NumConst} In our numerical calculation, we take
$\kappa=3$, $\mu=2\Delta_{\mathrm{so}}$, and $v_{F}=c/550$.
\bibitem{metallicSub}
In our simulations, the UPM's frequency is in the order
of $10$-$100$ meV, and within this region, one simple
calculation within classical electromagnetic theory gives
the same order amplitude in the absorption resonance of germanene
compared with the background spectra of the Ag substrate. Also,
when the absorption peak occurs at larger than $50$ meV, the
plasmon resonance would not be damped by the Ag substrate due
to the small relaxation time of Ag ($50$ fs corresponds to $12$ meV).
Furthermore, usually the energies of the surface and bulk plasmon
resonances of the metallic substrate lie in the order $1$ eV
(Ag is $3$ eV while Au is $2.2$ eV), and they are far away
from that of germanene, with the result that the interference
between the plasmon resonances of the substrate and germanene is
nearly impossible. All three factors make the plasmon
resonances of germanene observable in experiments, even in an
Ag substrate.
\end{thebibliography}
\end{document}